# A HIGGS-FREE MODEL FOR FUNDAMENTAL INTERACTIONS AND ITS IMPLICATIONS

**Marek Pawłowski**[1][†]
Soltan Institute for Nuclear Studies, Warsaw, POLAND
**and**
**Ryszard Rączka**[2][‡]
Soltan Institute for Nuclear Studies, Warsaw, POLAND
and
Interdisciplinary Laboratory for Natural and Humanistic Sciences
International School for Advanced Studies (SISSA), Trieste, ITALY

January, 1995

## Abstract

A model for strong, electroweak and gravitational interactions based on the local symmetry group $G = SU(3) \times SU(2)_L \times U(1) \times C$ where $C$ is the local conformal symmetry group is proposed. The natural minimal $G$-invariant form of total lagrangian is postulated. It contains all Standard Model fields and the gravitational interaction, however the Higgs mass term $\mu^2 \Phi^\dagger \Phi$ is forbidden. Using the unitary gauge and the conformal scale fixing conditions we can eliminate all four real components of the Higgs field in this model. In spite of that the tree level masses of vector mesons, leptons and quarks are automatically generated and are given by the same formulas as in the conventional Standard Model. In this manner one gets the mass generation without the mechanism of spontaneous symmetry breaking. We calculated in this model the predictions for a series of electroweak observables such as $m_{W_\pm}/m_Z$, $sin^2\theta_W^{eff}$, Z-boson widths, $A_{FB}^l$, etc, and we show that they are in agreement with experimental data. The gravitational sector of the model is also analyzed and it is shown that the model admits in the classical limit the Einsteinian form of gravitational interactions.

---
[1] Partially supported by Grant No. 2 P302 189 07 of Polish Committee for Scientific Researches.
[2] Partially supported by the Stiftung Für Deutsch-Polnische Zusammenarbeit Grant No. 984/94/LN.
[†] e–mail: PAWLOWSK@fuw.edu.pl
[‡] e–mail: RRACZKA@fuw.edu.pl

# 1 Introduction

The recent evidence for top quark production with the top mass estimated as $m_t = 174 \pm 10^{+13}_{-12} GeV$ [1] implies that the Higgs particle – if exists – may have the mass of the order of many hundred of $GeV$: in fact the central value of $m_H$ read off from the present central value of $m_t$ and electroweak (EW) data is $m_H \approx 300 GeV$ [2],[3]. It should be stressed however that some observables give very high central value for $m_H$: for instance using the value of $m_W$ as the input information one obtains that $m_H(m_W) \approx 1000 GeV$ with an enormous error however. Similar value of $m_H$ one can obtain from forward-backward asymmetry of $b\bar{b}$ pair production $A^b_{FB}$ (see Sec. 2). Some authors obtained even higher values of Higgs mass [4]. It is noteworthy that before the publication of the work [1] in most of electroweak calculations one assumed $m_H \approx 100 GeV$ whereas in the most recent works one uses in calculations $m_H \approx 300 GeV$ [2],[3],[5],[6]. Since in the lowest order $\lambda = \frac{1}{2}(\frac{m_H}{v})^2$ one can afraid that the Higgs self-coupling $\lambda$ would be also very large ($\lambda \approx 0.75$ for $m_H = 300 GeV$ and even $\lambda \approx 8$ for $m_H = 1000 GeV$). Such strong Higgs self-interaction would mean that the loops with Higgs particles would dominate all other contributions. Therefore the perturbative predictions in SM for many quantities become unreliable. Consequently the predictive power of the Standard Model (SM) and its consistency may be questionable.

The Higgs particle with such a large mass becomes suspicious. It is natural therefore to search for a modification of SM in which all confirmed by experiment particles would exist but the Higgs particle as the observed object would be absent.

We show in this work that such a modification of SM is possible under the condition that one joints to strong and electroweak interactions also the gravitational interaction. This extension of the class of SM interactions is in fact very natural. Indeed whenever we have the strong and electroweak interactions of elementary particles, nuclea, atoms or other objects we have also at the same time the gravitational interactions. It seems natural therefore to consider a unified model for strong, electroweak and gravitational interactions which would describe simultaneously all four fundamental interactions. It is well known that gravitational interactions give a negligible effect to most of strong or electroweak elementary particle processes. We show however that they may play the crucial role in a determination of the physical fields and their masses in the unified model and that their presence allows to eliminate all Higgs fields from the final lagrangian.

In turn we recall that in the conventional Standard Model the Higgs mechanism of spontaneous symmetry breaking (SSB) provides a simple and effective instrument for mass generation of weak gauge bosons, quarks and leptons. However, despite of many efforts of several groups of experimentalists [7] the postulated Higgs particle of the SM was not observed. Hence one might expect that the model for strong and electroweak interactions supplemented by the gravitational interaction in which all dynamical Higgs fields may be eliminated can provide a natural frame–work for a description of elementary particle fundamental interactions.

In order to construct a unique form of the theory of strong and electroweak interactions extended by the gravitational interactions we observe that the gauge symmetry $SU(3) \times SU(2)_L \times U(1)$ of the fundamental interactions may be naturally extended by the local conformal symmetry. The choice of the unitary gauge condition for $SU(2)_L$ gauge group allows to eliminate the three out of four real Higgs fields from the complex Higgs doublet. In turn the choice of the scale fixing condition connected with the local conformal symmetry allows to eliminate the last Higgs field. In that manner all four Higgs fields can be gauged away completely! It is remarkable that in spite of the elimination of all Higgs fields in our model the vector meson, lepton and quark masses are generated and at the tree level they are given by the same analytical formulas as in the conventional SM.

Thus it may be that the dynamical real Higgs field and the associated Higgs particles are in fact absent and it is therefore not surprising that they could not be detected in various experiments [7].

We review in Section 2 the present problems with a very massive Higgs particle. Next in Section 3 we discuss the properties of local conformal symmetry and its representations in field space of arbitrary spin. We present in Section 4 the form of the total lagrangian of our unified theory of electroweak, strong and gravitational interactions determined by the gauge and the local conformal invariance. The noteworthy feature of the obtained lagrangian is the lack of the Higgs mass term $\mu^2 \Phi^\dagger \Phi$. We show next that using the unitary gauge condition and the conformal scale fixing condition we can eliminate all dynamical Higgs fields from the theory! We show in Section 5 that in spite of the lack of dynamical Higgs fields the masses of vector mesons, leptons and quarks are generated and at the tree level they are given by the same analytical expressions in terms of coupling constants as in the conventional SM. We give in this section the path integral formulation of our model and show a remarkable result that conformal invariant products of fields have the conformal invariant vacuum expectation values.

We discuss in Section 6 the predictions of our model in electroweak sector. The elimination of all Higgs fields leads us in the flat space-time approximation to the model with massive vector mesons, which is nonrenormalizable. In order to get definite perturbative predictions – especially for electroweak processes – we have to introduce the ultraviolet cutoff $\Lambda$. We show the close connection between the large Higgs mass $m_H$ and $\Lambda$. We illustrate this relation in the case of universal electroweak parameters $\varepsilon_{N1}$, $\varepsilon_{N2}$ and $\varepsilon_{N3}$ introduced by Altarelli *et al.* [8]. We show that the difference between SM results for $\varepsilon_{Ni}$ and in our model is essentially proportional to $\log \frac{\Lambda^2}{m_H^2}$; thus if one chooses $\Lambda \cong m_H$ one obtains the same analytical formulas for $\varepsilon_{Ni}$ in SM and in our Higgs-free model up to the terms which vanish in the limit $m_H \to \infty$. Next we have calculated in one-loop approximation a series of electroweak observables such as $\Gamma_l$ – the lepton width of $Z$ mesons, the $m_W/m_Z$ ratio, the effective $sin^2\theta_W^{eff}$ of the Weinberg angle and others as the function of the UV cutoff $\Lambda$. Elimination of $\Lambda$ from these formulas leads to a relation between observables in our model. Taking $\Gamma_l$ as the "EW-meter" we have obtained the predictions for other observables which are in agreement with experimental data.

We remark also that using so called Generalized Equivalence Theorem one may calculate the high energy limit for various processes in our model.

We present in Section 7 the analysis of the gravitational sector in the unified model. We show that our unified model after determination of the unitary gauge and scale fixing leads already on the classical level to the conventional gravitational theory with Einstein–Hilbert lagrangian implied by the conformal Penrose term contained in the unified lagrangian.

Finally we discuss in Section 8 several basic problems connected with a description of fundamental interactions which are given by the conventional SM or its extensions and by our nonrenormalizable Higgs-free model. We discuss also some open problems connected with derivation of predictions in low and high energy regions from nonrenormalizable Higgs-free models.

The present work is the extension of our two previous papers [9],[10] and contains the answer to several questions raised by theirs readers.

## 2  Difficulties with Standard Model Higgs particle.

We shall argue that the recently announced [1] evidence for the top quark with the mass

$$m_t = 174 \pm 10^{+13}_{-12} GeV \qquad (2.1)$$

may lead to a serious conceptual and calculational problems in the Standard Model. The relatively heavy top quark with the mass (2.1) – heavier than expected on the base of LEP1-CDF-UA1 data [5],[11]-[13] – shifts up the expected region of SM Higgs mass and consequently also the area of expected Higgs quartic self-coupling $\lambda$.

We present, for an illustration, the central values of $m_H$ from various observables. Setting the central values $m_t = 174 GeV$, $\alpha(m_Z) = 1/128.87$, $\alpha_s = 0.123$, and $m_Z = 91.1888 GeV$ [3],[6] one obtains within the minimal SM the electroweak observables as the functions of $m_H$. We give in Fig. 2.1 the plot of various EW observables in dependence of $m_H$ calculated by means of the newest version of the code ZFITTER [14] (v.4.8 of 07.09.94).

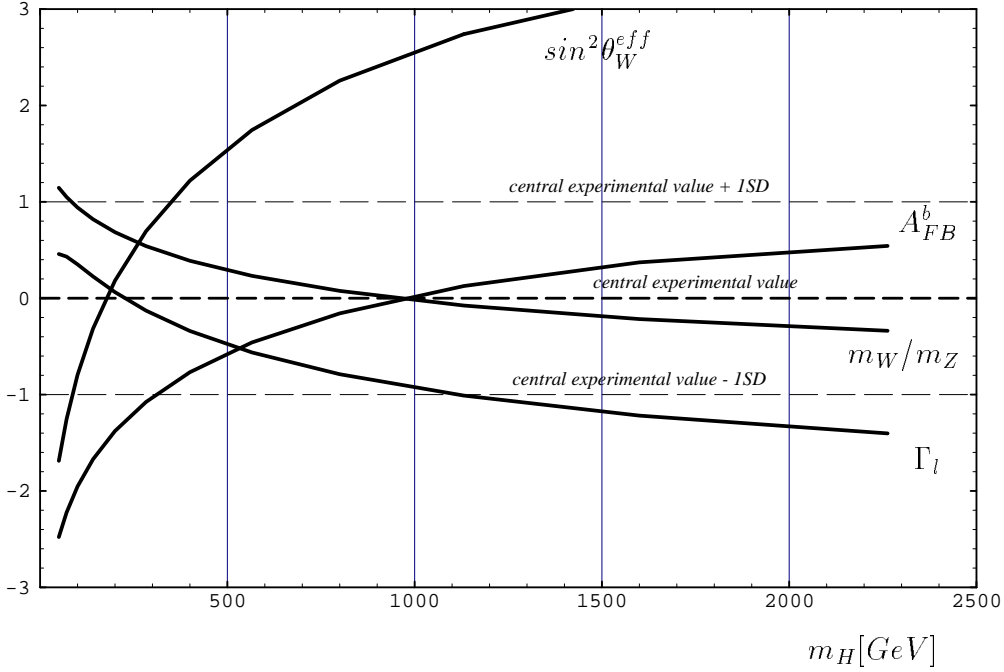

**Fig. 2.1** *The SM predictions for the dependence of various EW observables on $m_H$ for $m_t$ fixed at 174GeV. Each of the observables is shifted down by its central experimental value and is rescaled by its one standard deviation. Consequently the central values (thick dashed line) and the standard deviations (thin dashed lines) of various observables are situated at the same place of the plot.*

The central values of $sin^2\theta_W^{eff}$, $\Gamma_l$, $m_W/m_Z$ and $A_{FB}^b$ imply that the central values of $m_H$ are

$$m_H(sin^2\theta_W^{eff}) \approx m_H(\Gamma_l) \approx 200 GeV$$
$$m_H(A_{FB}^b) \approx m_H(m_W/m_Z) \approx 1000 GeV \qquad (2.2)$$

where we have taken $m_W = 80.23 \pm 0.18 GeV$ [15] what implies that $m_W/m_Z = 0.8798 \pm 0.0020$.

Since the Higgs self–coupling constant $\lambda$ and the Higgs mass are connected at the tree level by the formula

$$\lambda = \frac{1}{2}(\frac{m_H}{<\phi>})^2, \qquad <\phi> = 246 GeV \tag{2.3}$$

one obtains

$$\lambda(sin^2\theta_W^{eff}) \approx \lambda(\Gamma_l) \approx \frac{1}{3}$$
$$\lambda(A_{FB}^b) \approx \lambda(m_W/m_Z) \approx 8. \tag{2.4}$$

This looks rather dangerous; however to be honest we should conclude from Fig. 2.1 that within the present experimental errors there is a considerable admissible deviation from the values given by (2.2) [5],[12]. Consequently smaller values of $m_H$ and therefore also smaller values of $\lambda$ are not excluded.

Despite the fact that the present electroweak data are not very conclusive the result (2.1) compels many authors to consider the possibilities of large Higgs mass and strong Higgs self–coupling more seriously [4]. In fact in most of the recent analysis of electroweak data one assumes $m_H \approx 300 GeV$ instead of the value $m_H \approx 100 GeV$ in previous analysis and many authors consider the models in the limit $m_H \to \infty$ [16].

The rather strong Higgs self–coupling like (2.4) may break–down the perturbative calculations for many processes for which Higgs loops with $\lambda$-coupling contributes. For instance the two-loop perturbation expansion for the partial width $\Gamma(H \to \bar{f}f)$ of the Higgs particle decay into the fermion – anti-fermion pair can be written in the form

$$\Gamma(H \to \bar{f}f) = \Gamma_0[1 + 0.11(\frac{m_H}{1TeV})^2 - 0.78(\frac{m_H}{1TeV})^4] \tag{2.5}$$

where $\Gamma_0$ is the partial width in the Born approximation and the second and third term in the bracket represent the one- and the two-loop contributions respectively [17].

We see that with increasing $m_H$ the importance of the two-loop contribution rapidly increases: in fact for $m_H > 375 GeV$ the two-loop contribution dominates the one-loop and for $m_H > 1200 GeV$ the width becomes negative! This demonstrates the complete breakdown of perturbation theory for the Higgs mass of the order of 1TeV.

We see therefore that the supposition that the real Higgs field and the corresponding Higgs particle exists in the Standard Model may lead to rather fundamental conceptual and calculational difficulties. Therefore it seems justified at present to look for a modification of SM in which all experimentally confirmed facts would be reproduced but the Higgs particle as the observed object would not exist.

The Higgs sector and the Higgs mechanism of mass generation looked suspicious to many physicists since the beginning of its introduction. In fact Kuminasa and Goto already in 1967 have proposed a Higgs-free model of gauge field theory for massive vector mesons interacting with fermions [18]. Next the Higgs-free models for electroweak interactions were considered from various points of view [19].

Recently there were proposed several new Higgs-free models for electroweak and strong interactions. In particular Schildknecht and collaborators proposed the Higgs-free massive vector boson model [20] and they have compared some of its predictions with the predictions of the conventional SM. In the work [21] it was proposed a Higgs-free SM with nonrenormalizable current–current and dipole–dipole interactions. The EW models with boson condensates were proposed by several authors [22]. Finally in [16] it was proposed a gauged $\sigma$-model for electroweak interactions.

It seems to us that our Higgs-free model based on the extension of electroweak and strong interactions by gravitational interactions, which leads to the extension of gauge symmetry by the local conformal symmetry, presents a most natural frame–work for a description of fundamental interactions.

## 3 Local conformal symmetry

Let $M^{3,1}$ be the pseudo–Riemannian space time with the metric $g_{\alpha\beta}$ with the signature $(+, -, -, -)$. Let $\Omega(x)$ be a strictly positive function on $M^{3,1}$ which has the inverse $\Omega^{-1}(x)$. Then the local conformal transformation in $M^{3,1}$ is defined as the transformation which changes the metric by the formula

$$g_{\mu\nu}(x) \to \tilde{g}_{\mu\nu}(x) = \Omega^2(x) g_{\mu\nu}(x). \qquad (3.1)$$

The set of all local conformal transformations forms the multiplicative abelian infinite–dimensional group $C$ with the obvious group multiplication law.

It is evident from (3.1) that $(M^{3,1}, g_{\mu\nu})$ and $(M^{3,1}, \tilde{g}_{\mu\nu})$ have identical causal structure and conversely it is easy to show that any two space times which have identical causal structure must be related by a local conformal transformation.

The conformal transformations occur in many problems in general relativity. In particular Canuto et. al. proposed the scale–covariant theory of gravitation, which provides an interesting alternative for the conventional Einstein theory [23].

It should be stressed that a conformal transformation is not a diffeomorphism of space time. The physical meaning of the conformal transformations follows from the transformation law of the length element

$$dl(x) = \sqrt{-g_{ij}dx^i dx^j} \qquad \rightarrow \qquad d\tilde{l}(x) = \Omega(x)dl(x). \qquad (3.2)$$

Hence a local conformal transformation changes locally the length scale. Since in some places of the Earth one utilizes *the meter* as the length scale, whereas in other places one utilizes *the feet* or *the ell* as the length scale one my say that one utilizes the local conformal transformations in everyday live. Similarly one verifies that the conformal transformation changes locally the proper time

$$ds(x) = \sqrt{g_{\mu\nu}dx^\mu dx^\nu} \qquad \rightarrow \qquad d\tilde{s}(x) = \Omega(x)ds(x).$$

Since the physical phenomena should be independent of the unit chosen locally for the length, the proper time, mass etc. the group $C$ of local conformal transformations should be a symmetry group of physical laws.

In order to avoid any confusion we stress that the abelian group $C$ has nothing in common with the 15 parameter nonabelian conformal group $SO(4,2)$ defined locally in the $M^{3,1}$ by the action of Poincare, dilatation and special conformal transformations. It is remarkable however that the gauge theory based on $SO(4,2)$ is equivalent to conformal gravity implied by $C$-invariance [24].

Comparing the physical meaning of local conformal transformations and the local gauge $SU(2)_L$ transformations of SM associated with the concept of the weak isospin it seems that the conformal transformations are not less natural symmetry transformation than the nonabelian gauge transformations in the SM.

We shall give now a construction of the representation of the conformal group $C$ in the field space. Let $\Psi$ be a tensor or spinor field of arbitrary spin. Define the map

$$\Omega \rightarrow U(\Omega)$$

by the formula

$$\tilde{\Psi}(x) = U(\Omega)\Psi(x) = \Omega^s(x)\Psi(x), \qquad s \in R \qquad (3.3)$$

The number $s$ is determined by the condition of conformal invariance of field equation. We say that field equation for $\Psi$ is conformal invariant if there exist $s \in R$ such that $\Psi(x)$ is a solution with the metric $g_{\mu\nu}(x)$ if and only if $\tilde{\Psi}(x)$ given by (3.3) is a solution with the metric $\tilde{g}_{\mu\nu}(x)$. The number $s$ is called the conformal weight of $\Psi$ [25], [26], [27]. It is evident that the map $\Omega \to U(\Omega)$ defines the representation of $C$ in the field space.

Using the above definitions one can calculate the conformal weight for a field of arbitrary spin. One finds for instance that the Maxwell field $F_{\mu\nu}$ on $(M^{3,1}, g)$ has the conformal weight $s = 0$ whereas $F^{\mu\nu}$ has $s = -4$.

Similarly one can show that the Yang–Mills field strength $F_{\mu\nu}{}^a$ has the conformal weight $s = 0$ whereas the massless Dirac field has the conformal weight $s = -\frac{3}{2}$. It is noteworthy that the scalar massless field $\Phi$ satisfying the Laplace–Beltrami equation

$$\triangle \Phi = 0$$

is not conformal invariant. In fact it was discovered by Penrose that one has to add to the Lagrangian on $(M^{3,1}, g)$ the term

$$-\frac{1}{6} R \Phi^\dagger \Phi$$

where $R$ is the Ricci scalar, in order that the corresponding field equation is conformal invariant with the conformal weight $s = -1$ [28].

# 4 A unified model for strong, electroweak and gravitational interactions

We postulate that the searched unified theory of strong, electroweak and gravitational interactions will be determined by the condition of invariance with respect to the group $G$

$$G = SU(3) \times SU(2)_L \times U(1) \times C \tag{4.1}$$

where $C$ is the local conformal group defined by (3.1). Let $\Psi$ be the collection of vector meson, fermion and scalar fields which appear in the conventional minimal SM for electroweak and strong interactions. Then the minimal natural conformal and $SU(3) \times SU(2)_L \times U(1)$ –gauge invariant total lagrangian $L(\Psi)$ may be postulated in the form:

$$L = [L_G + L_F + L_Y + L_\Phi + L_{grav}]\sqrt{-g} \tag{4.2}$$

Here $L_G$ is the total lagrangian for the gauge fields $A_\mu^a$, $W_\mu^b$ and $B_\mu$, $a = 1, ..., 8$, $b = 1, 2, 3$ associated with $SU(3) \times SU(2)_L \times U(1)$ gauge group

$$L_G = -\frac{1}{4} F^a{}_{\mu\nu} F^{a\,\mu\nu} - \frac{1}{4} W^b{}_{\mu\nu} W^{b\,\mu\nu} - \frac{1}{4} B_{\mu\nu} B^{\mu\nu}, \tag{4.3}$$

and $F^a{}_{\mu\nu}$, $W^b{}_{\mu\nu}$ and $B_{\mu\nu}$ are the conventional field strengths of gauge fields in which the ordinary derivatives are replaced by the covariant derivatives e.g.

$$B_{\mu\nu} = \nabla_\mu B_\nu - \nabla_\nu B_\mu, \qquad (4.4)$$

etc.; $L_F$ is the lagrangian for fermion field interacting with the gauge fields; $L_Y$ represents the Yukawa interactions of fermion and scalar fields; $L_\Phi$ is the $G$-invariant lagrangian for the scalar fields, which may be written in the form:

$$L_\Phi = (D\Phi)^\dagger (D\Phi) - \lambda(\Phi^\dagger \Phi)^2 + \beta \partial_\mu |\Phi| \partial^\mu |\Phi| - \frac{1}{6}(1+\beta)R\Phi^\dagger \Phi, \qquad (4.5)$$

where $D$ denotes the covariant derivative with connections of all symmetry groups. Notice that the condition of conformal invariance does not admit the Higgs mass term $\mu^2 \Phi^\dagger \Phi$ which assures the mechanism of spontaneous symmetry breaking and mass generation in the conventional formulation. Instead we have two additional terms: the Penrose term

$$-\frac{1}{6}(1+\beta)R\Phi^\dagger \Phi \qquad (4.6)$$

which assures that the lagrangian (4.5) is conformal invariant, and the term

$$\beta \partial_\mu |\Phi| \partial^\mu |\Phi| \qquad (4.7)$$

which together with the term $-\frac{1}{6}\beta R \Phi^\dagger \Phi$ is conformal and gauge invariant. It may be surprising that (4.7) depends on $|\Phi|$. Observe however that the conventional first term in $L_\Phi$ can be written in the form

$$(D\Phi)^\dagger (D\Phi) = \partial_\mu |\Phi| \partial^\mu |\Phi| + |\Phi|^2 L_\sigma(g(\Phi), W, B) \qquad (4.8)$$

where $g(\Phi)$ is $SU(2)_L$ gauge unitary matrix defined by the formula

$$\Phi = \begin{pmatrix} \phi_u \\ \phi_d \end{pmatrix} = g(\Phi) \begin{pmatrix} 0 \\ |\Phi| \end{pmatrix}, \quad g(\Phi) = \frac{1}{|\Phi|} \begin{pmatrix} \bar{\phi}_d & \phi_u \\ -\bar{\phi}_u & \phi_d \end{pmatrix} \qquad (4.9)$$

and $L_\sigma(g(\Phi), W, B)$ is a gauged–sigma–model–like lagrangian.

We see therefore that the term like (4.7) is already present in the conventional gauge invariant lagrangian.

The last term in (4.2) is the Weyl term

$$L_{grav} = -\rho C^2, \qquad \rho > 0, \qquad (4.10)$$

where $C^\delta_{\alpha\beta\gamma}$ is the Weyl tensor which is conformally invariant. Using the Gauss–Bonnet identity we can write $C^2$ in the form

$$C^2 = 2(R^{\mu\nu}R_{\mu\nu} - \frac{1}{3}R^2). \tag{4.11}$$

We see that the condition of conformal invariance does not admit in (4.2) the conventional gravitational Einstein lagrangian

$$L = \kappa^{-2}R\sqrt{-g}, \qquad \kappa^2 = 16\pi G. \tag{4.12}$$

It was shown however by Stelle [29] that quantum gravity sector contained in (4.2) is perturbatively renormalizable whereas the quantum gravity defined by the Einstein lagrangian (4.12) coupled with matter is nonrenormalizable [30]. Hence, for a time being it is an open question which form of gravitational interaction is more proper on the quantum level. We show in Section 7 that the Einstein lagrangian (4.12) may be reproduced by Penrose term if the physical scale is properly determined. The discussion of the role of quantum effects which may reproduce the lagrangian (4.12) and give the classical Einstein theory as the effective induced gravity was presented in our previous work [10].

Notice that conformal symmetry implies that all coupling constants in the present model are dimensionless.

The theory given by (4.2) is our conformally invariant proposition alternative to the standard Higgs–like theory with SSB. Its new, most important feature is the local conformal invariance. It means that simultaneous rescaling of all fields (including the field of metric tensor) with a common, arbitrary, space–time dependent factor $\Omega(x)$ taken with a proper power for each field (the conformal weight) will leave the Lagrangian (4.2) unaffected. The symmetry has a clear and obvious physical meaning [31], [26]. It changes in every point of the space–time all dimensional quantities (lengths, masses, energy levels, etc) leaving theirs ratios unchanged. It reflexes the deep truth of the nature that nothing except the numbers has an independent physical meaning.

The freedom of choice of the length scale is nothing but the scale fixing freedom connected with the conformal symmetry group. In the conventional approach we define the length scale in such a way that elementary particle masses are the same for all times and in all places. This will be the case when we rescale all fields with the $x$–dependent conformal factor $\Omega(x)$ in such a manner that the length of the rescaled scalar field doublet is fixed i.e.

$$\tilde{\Phi}^\dagger \tilde{\Phi} = \frac{v^2}{2} = const. \tag{4.13}$$

(We shall discuss the problem of mass generation in details in Section 5.)

The scale fixing for the conformal group (4.13) is distinguished by nothing but our convenience. Obviously we can choose other scale fixing condition, e.g. we can use the freedom of conformal factor to set

$$\sqrt{-\tilde{g}} = 1; \qquad (4.14)$$

this will lead to other local scales but as we show below it will leave physical predictions unchanged.

Our model with the scale fixing condition (4.13) considered in the flat space-time limit coincides with the gauged nonlinear $\sigma$-model analyzed in several recent papers [16]. Hence all results obtained for this model are applicable also in our model.

It follows from Fadeev-Popov method that the expectation values of gauge invariant operators are gauge invariant i.e. they are independent on a chosen gauge fixing condition. We shall derive now the analogous result for the local conformal group and show that the expectation values of conformal invariant operators are independent on the choice of scale fixing condition.

In order to show this we shall use the functional integral formalism. Let $L[\Psi]$ be the scale invariant lagrangian (4.2). Let $C(\Psi)$ be the function of field operators which is local conformal invariant i.e.

$$C(\tilde{\Psi}) = C(\Psi)$$

where $\tilde{\Psi} = \Omega^{s_\Psi} \Psi$ is the conformal transform of scalar, vector or fermion field respectively given by (3.3) and determined by they conformal degree $s_\Psi$. Then according to the so called Matthews theorem the path integral representation for vacuum expectation values of $C(\Psi)$ has the form [32]:

$$< C(\Psi) >_0 = Z^{-1} \int C(\Psi) e^{iS_T(\Psi)} \Delta_f(\Psi) \delta[f(\Psi)] D\Psi \qquad (4.15)$$

where $Z$ is the partition function

$$Z = \int e^{iS_T(\Psi)} \Delta_f(\Psi) \delta[f(\Psi)] D\Psi \qquad (4.16)$$

$S_T = S + S_{FP}$ where $S_{FP}$ is the Fadeev-Popov contribution to the action integral due to the gauge fixing conditions and $f(\Psi)$ is the scale fixing condition. $D\Psi$ is the functional measure over all dynamical fields in $\Psi$ and in our case has the form

$$D\Psi = D\Phi DAD\psi Dg \qquad (4.17)$$

We chose the gauge fixing condition in such a manner that $S_{FP}$ is conformal invariant. It follows from Fadeev-Popov formalism [33] that

$$\Delta_f(\Psi) \int \delta[f(\Psi^\Omega)] D\Omega = I \tag{4.18}$$

where $D\Omega$ is the invariant measure on the conformal group and is given by the formula

$$D\Omega = \prod_x \frac{d\Omega(x)}{\Omega(x)} \tag{4.19}$$

One readily verifies that this measure is invariant under the group multiplication $\Omega \to \Omega'\Omega$ and the inversion $\Omega \to \Omega^{-1}$.

It follows from the conformal invariance of $D\Omega$ that $\Delta_f(\Psi)$ is conformal invariant. Setting as in (4.13)

$$f(\Psi^\Omega) = |\Phi^\Omega| - \frac{v}{\sqrt{2}}$$

and using the measure invariance we obtain

$$\Delta_f(\Psi) = \frac{v}{\sqrt{2}}$$

We present now the important result:
**Theorem 4.1**
Let $C(\Psi)$ be the conformal invariant function of field operators. Then the vacuum expectation value $< C(\Psi) >_0$ given by (4.15) is independent on the scale fixing condition.

(For the proof see Appendix.)

This result is a little bit surprising, especially if one takes into account how different are the scale fixing conditions (4.13) and (4.14). Theorem 4.1 implies that we can calculate the vacuum expectation values of conformal invariant function of field operators using the most convenient scale fixing condition. Since the condition (4.13) together with the unitary gauge fixing condition for $SU(2)_L$ group eliminates all four Higgs fields from the action integral $S_T(\Psi)$ we shall use it exclusively in all following calculations. We note that the scattering operator $\hat{S}$ is dimensionless and therefore conformal invariant. Consequently if we use the normalization of asymptotic states such that they are dimensionless we can use the scale fixing condition (4.13) for calculation of probability amplitudes of all physical processes.

# 5 Generation of lepton, quark and vector boson masses

We demonstrate now that using the conformal group scale fixing condition (4.13) we can generate the same lepton, quark and vector meson masses as in the conventional SM without however use of any kind of Higgs mechanism and SSB.

In fact inserting the scale fixing condition (4.13) into the Lagrangian (4.2) we obtain

$$\tilde{L} = L^{scaled} = [L_G + L_F + L_Y^{scaled} + L_\Phi^{scaled} + L_{grav}]\sqrt{-g}, \qquad (5.1)$$

in which the condition (4.13) was inserted into $L_\Phi$ and $L_Y$. We should use the symbol $\tilde{\Phi}$, $\tilde{\Psi}$ etc. for the rescaled fields in (5.1), however for the sake of simplicity we shall omit "~" sign over fields in the following considerations.

The condition (4.13) together with the unitary gauge fixing of $SU(2)_L \times U(1)$ gauge group, reduce by (4.9) the Higgs doublet to the form

$$\Phi^{scaled} = \frac{1}{\sqrt{2}}\begin{pmatrix} 0 \\ v \end{pmatrix}, \qquad v > 0 \qquad (5.2)$$

and produce the tree level mass terms for leptons, quarks and vector bosons associated with $SU(2)_L$ gauge group. For instance the $\Phi$–lepton Yukawa interaction $L_Y^l$ reads

$$L_Y^l = -\sum_{i=e,\mu,\tau} G_i \bar{l}_{iR}(\Phi^\dagger l_{iL}) + h.c.$$

where

$$l_{eL} = \begin{pmatrix} \nu_e \\ e_L \end{pmatrix} \quad etc.$$

It passes into

$$L_Y^{l\ scaled} = -\frac{1}{\sqrt{2}}v(G_e \bar{e}e + G_\mu \bar{\mu}\mu + G_\tau \bar{\tau}\tau) \qquad (5.3)$$

giving the conventional, space–time independent lepton masses

$$m_e = \frac{1}{\sqrt{2}}G_e v, \qquad m_\mu = \frac{1}{\sqrt{2}}G_\mu v, \qquad m_\tau = \frac{1}{\sqrt{2}}G_\tau v. \qquad (5.4)$$

Similarly one generates from $\Phi$–quark Yukawa interaction $L_Y^q$ the corresponding quark masses. In turn from $L_\Phi$-lagrangian (4.5) using the scaled scalar field (5.2) one obtains

$$(D_\mu \Phi)^\dagger D^\mu \Phi = \frac{g_2^2 v^2}{4} W_\mu^+ W^{\mu-} + \frac{g_1^2 + g_2^2}{8} v^2 Z^2$$

where

$$Z_\mu = -\sin\theta_W B_\mu + \cos\theta_W W^3{}_\mu, \qquad \cos\theta_W = \frac{g_2}{\sqrt{g_1^2 + g_2^2}}.$$

Hence one obtains the following vector mesons masses

$$m_W = \frac{v}{2} g_2, \qquad m_Z = \frac{m_W}{\cos\theta_W}. \tag{5.5}$$

It is remarkable that the analytical form for tree level fermion and vector meson masses in terms of coupling constants and the parameter $v$ is the same as in the conventional SM. We see therefore that the Higgs mechanism and SSB is not indispensable for the fermion and vector mesons mass generation!

We note that the fermion–vector boson interactions in our model are the same as in SM. Hence analogously as in the case of conventional formulation of SM one can deduce the tree level relation between $v$ and $G_F$ – the four–fermion coupling constant of $\beta$–decay:

$$v^2 = (2G_F)^{-1} \to v = 246 GeV. \tag{5.6}$$

Here we have used the standard decomposition $g^{\mu\nu}\sqrt{-g} = \eta^{\mu\nu} + \kappa' h^{\mu\nu}$ (see e.g. [34]) which reduces the tree level problem for the matter fields to the ordinary flat case task.

We see therefore that the resulting expressions for masses of physical particles are identical as in the conventional SM.

Let us stress that the scale fixing condition like (4.13) does not break $SU(2)_L \times U(1)$ gauge symmetry. The symmetry is broken (or rather one of gauge equivalent description is fixed) when (4.13) is combined with unitary gauge condition of electroweak group leading to (5.2). However, also after imposing of a gauge condition like (5.2) we have a remnant of both the conformal and $SU(3) \times SU(2)_L \times U(1)$ initial gauge symmetries: this is reflected in the special, unique relations between couplings and masses in our model

# 6  Precision tests of electroweak interactions.

Our model represents in fact the gauge field theory model with massive vector mesons and fermions. It is well–known that such models are in general nonrenormalizable [35],[36]. We remind however that in the nonrenormalizable Fermi model for weak interactions we can make a definite predictions for low energy phenomena e.g. for $\mu$ or neutron decays. Similarly the recent progress with so called Generalized Equivalence Theorem allows to make definite predictions for the scattering operator in nonrenormalizable models like gauged nonlinear $\sigma$–model or other nonrenormalizable gauge field theory models [37]. Hence in our model we can obtain definite predictions for electroweak phenomena if we consider processes with energy $\sqrt{s}$ below some ultraviolet (UV) cutoff $\Lambda$. We wish to demonstrate that the cutoff $\Lambda$ is closely connected with the Higgs mass $m_H$ appearing in the Standard Model. Hence, from this point of view, Higgs mass is nothing else as the UV cutoff which assures that the truncated perturbation series is meaningful. We shall try to elucidate this problem on the example of so called precision tests of electroweak theory.

One–loop radiative corrections to various electroweak quantities or processes can be expressed in terms of three quantities $\Delta r$, $\Delta \rho$ and $\Delta k'$. We refer to the recent excellent reviews for the precise definitions of these quantities and for their analytical expressions [3],[38][39]. For an illustration we recall that the expression for W–meson mass, up to one loop order, has the form

$$m_W = \frac{m_Z}{\sqrt{2}} \left\{ 1 + \sqrt{1 + \frac{2\sqrt{2}\pi\alpha}{m_Z G_F (1-\Delta r)}} \right\}^{\frac{1}{2}} \qquad (6.1)$$

where $\Delta r(m_t, m_H)$ is the one loop correction to $\mu$–decay amplitude which in Standard Model depends on top and Higgs masses.

It was suggested by Altarelli *et.al* [8] to pass from $\Delta r$, $\Delta \rho$ and $\Delta k'$ to new quantities $\varepsilon_{N1}$, $\varepsilon_{N2}$ and $\varepsilon_{N3}$ such that $\varepsilon_{N2}$ and $\varepsilon_{N3}$ depend on $m_t$ only logarithmically. These parameters characterize the degree of $SU(2)_L \times U(1)$ symmetry breaking and their numerical value significantly different from zero would signal a "new physics" [8],[20].

If we calculate these parameters in our model in one–loop approximation we find the specific class of Feynman diagrams with fermion and vector boson loops which contributes to them. Since some vector boson loops will produce divergences, e.g. in the case of fermion – massive vector boson coupling constant, one has to introduce either the new renormalization constants or UV cutoff $\Lambda$ which can be given by the formula [20]

$$\log \frac{\Lambda^2}{\mu^2} = \frac{2}{4-D} - \gamma_E + \log 4\pi + \frac{5}{6} \qquad (6.2)$$

where $\mu$ is the reference mass of dimensional regularization, $D$ is the space–time dimension and $\gamma_E$ is the Euler's constant.

One obtains the formula for $\varepsilon_{Ni}$ parameters in SM if one adds to the class of Feynman diagrams in our model all appropriate one–loop diagrams with Higgs internal lines. Using the results of [20] and [40] one obtains

$$\varepsilon_{N_1}^{SM} - \varepsilon_{N_1}^{HFM} = \frac{3\alpha(m_Z^2)}{16\pi c_0^2} \log\left(\frac{\Lambda^2}{m_H^2}\right) + \varepsilon_{N_1 rem}(X)$$

$$\varepsilon_{N_2}^{SM} - \varepsilon_{N_2}^{HFM} = \varepsilon_{N_2 rem}(X) \quad (6.3)$$

$$\varepsilon_{N_3}^{SM} - \varepsilon_{N_3}^{HFM} = \frac{\alpha(m_Z^2)}{48\pi s_0^2} \log\left(\frac{\Lambda^2}{m_H^2}\right) + \varepsilon_{N_3 rem}(X)$$

$$X = \frac{m_Z^2}{m_H^2} \log \frac{m_Z^2}{m_H^2}$$

where $HFM$ index of $\varepsilon_{N_i}$ means that the quantity was calculated in our Higgs-Free Model. Here $\alpha(m_Z^2) = \frac{1}{129}$ and $c_0$ and $s_0$ are defined by the formula

$$s_0^2(1 - s_0^2) = s_0^2 c_0^2 \equiv \frac{\pi\alpha(m_Z^2)}{\sqrt{2} G_F m_Z^2}$$

.

The above formulas indicate a role which plays in SM the very large Higgs mass: first the numerical analysis shows that the term $\varepsilon_{N_i rem}(X)$ for $m_H \geq 300 GeV$ can be disregarded and second if we take the UV cutoff $\Lambda \simeq m_H$ then by (6.3) the prediction for $\varepsilon_{Ni}$–parameters in the conventional SM and our nonrenormalizable model almost coincide. Thus the very large Higgs mass preferred by the top mass $m_t = 174 GeV$ plays in fact in the conventional SM the role of UV cutoff parameter. If the Higgs particle will be not found then our model provides an extremely natural frame–work for the description of electroweak and strong interactions at least up to TeV energies.

We would like to discuss now the problem of getting predictions from our nonrenormalizable model for quantities like $W$-meson mass $m_W$, $sin^2\theta_W^{eff}$ of effective Weinberg angle, lepton width $\Gamma_l$ and other characteristics of $Z$ peak in $e^+ e^-$ collision which are measured in so called precision tests of electroweak theories. It is known that the SM predictions for these quantities including one-loop radiative corrections depend on the unknown value of the Higgs mass. On the contrary, the one loop-predictions of our model depend on the cutoff parameter $\Lambda$. One can calculate these predictions directly or one can use SM results and correct them using (6.3). The expressions for $\varepsilon_{N_i rem}(X)$ are known and are given explicitly in [40].

We give in Fig. 6.1 the plot of $\Gamma_l$ as the function of $\Lambda$.

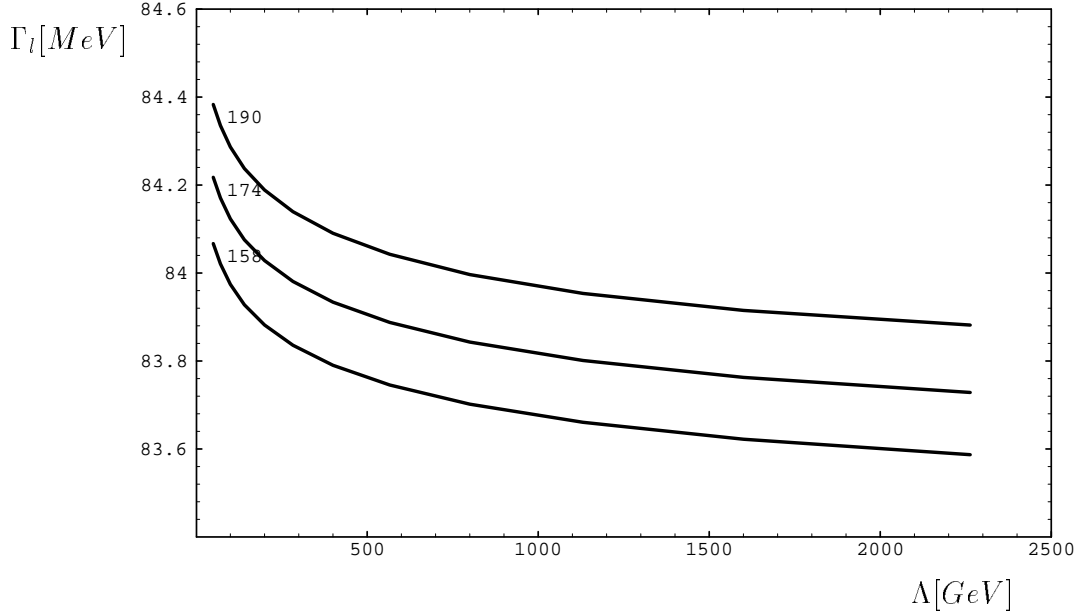

**Fig. 6.1** *The plot of $\Gamma_l$ as the function of UV cutoff $\Lambda$ predicted within Higgs-free model for top masses $m_t = 158 GeV$, $m_t = 174 GeV$ and $m_t = 190 GeV$.*

Similarly we can plot the corresponding figures for every other measurable quantity from the considered set.

All our one-loop predictions will be $\Lambda$-cutoff dependent and the precise value of the $\Lambda$-cutoff is unknown. In order to obtain the definite predictions we have to select some "EW-meter" i.e. a quantity $R_0(\Lambda)$ which is measured with the best accuracy in the present EW experiments and which will replace unknown $\Lambda$ in the expressions for the other physical quantities $R_i$. Inverting the relation $R_0(\Lambda)$ we can express $\Lambda$ as the function of $R_0$: $\Lambda(R_0)$. Then we insert this relation into the expression for any other quantity $R_i$ and we get the cutoff-independent definite function

$$R_i = \tilde{R}_i(R_0) = R_i(\Lambda(R_0)) \tag{6.4}$$

as the prediction of the model.

We assume that the best candidate for "EW-meter" must fulfill the following criteria:

i) should be measured directly (what excludes $sin^2\theta_W^{eff}$ obtained combining results of different asymmetries),

ii) should be of purely electroweak character at one-loop level (what excludes $b$-pair asymmetries) and

iii) should be measured with best accuracy relatively to the slope of its $\Lambda$ dependence what means that the ratio

$$\frac{\Delta_{exp}R}{dR/d\Lambda} \tag{6.5}$$

must be minimal at the measured central value of $R$.

The numerical analysis indicates that the "EW-meter" is presently given by $\Gamma_l$ observable.

We calculated within our model the $\Gamma_l$-dependence (6.4) for several most characteristic quantities measured in the precision tests of EW theories.

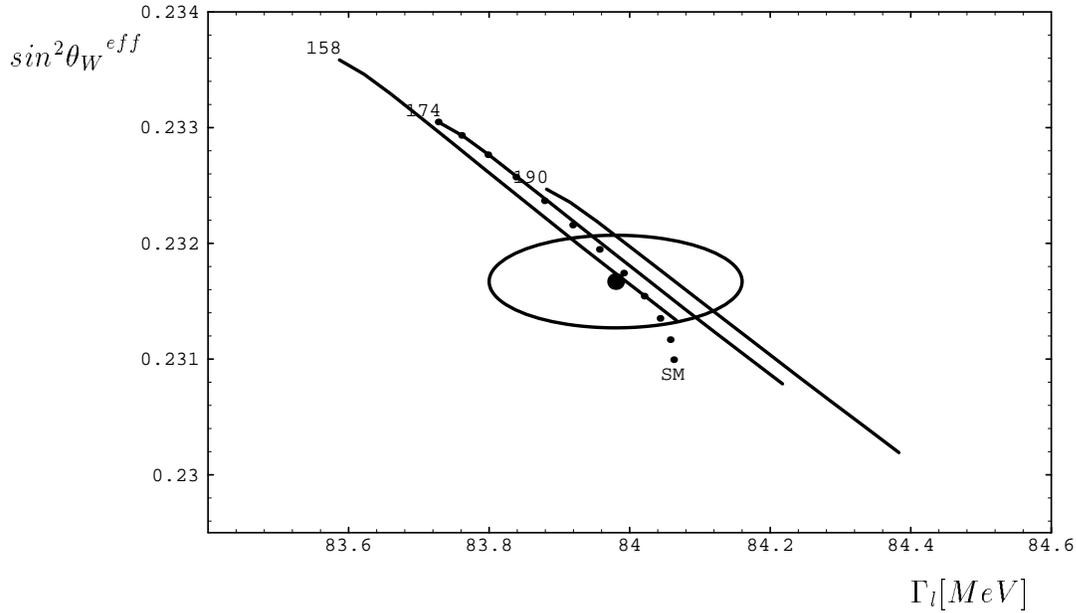

**Fig. 6.2** *The plot of $sin^2\theta_W^{eff}$ as the function of our "EW-meter" $\Gamma_l$ for top masses $m_t = 158 GeV$, $m_t = 174 GeV$ and $m_t = 190 GeV$. The dotted curve represents SM prediction for $m_t = 174 GeV$ and $m_H$ varying logarithmically from 50GeV (right side of the curve) to 2.25TeV. The circle is plotted at the central experimental value and the ellipse is plotted at 1 standard deviation.*

We give in Fig. 6.2 the plot of the ratio $sin^2\theta_W^{eff}$ as the function of $\Gamma_l$ obtained in our model after elimination of the ultraviolet cutoff $\Lambda$.

The three continuous curves correspond to our predictions for $sin^2\theta_W^{eff}$ as the function of $\Gamma_l$ for various top masses. The ball represents the central experimental values for $sin^2\theta_W^{eff}$ and $\Gamma_l$ and the ellipse is plotted at one standard deviation from this values. The dotted curve represents the SM predictions for $m_t = 174 GeV$. We see that $\Lambda$ cutoff independent predictions from our model agree surprisingly well with experimental data. In fact taking into account that our "EW-meter" $\Gamma_l$ is given by the experiment as

$$\Gamma_l = 83.98 \pm 0.18 MeV \tag{6.6}$$

we get our model prediction (for $m_t = 174 GeV$)

$$(sin^2\theta_W^{eff})^{HFM} = 0.2318 \pm .0008.$$

This should be confronted with the experimental value [6]

$$(sin^2\theta_W^{eff})^{EXP} = 0.23167 \pm 0.0004$$

and with SM prediction [3]

$$(sin^2\theta_W^{eff})^{SM} = 0.2322 \pm 0.0003 \pm 0.0006.$$

We give in Fig. 6.3 the plot of $m_W/m_Z$ as the function of $\Gamma_l$.

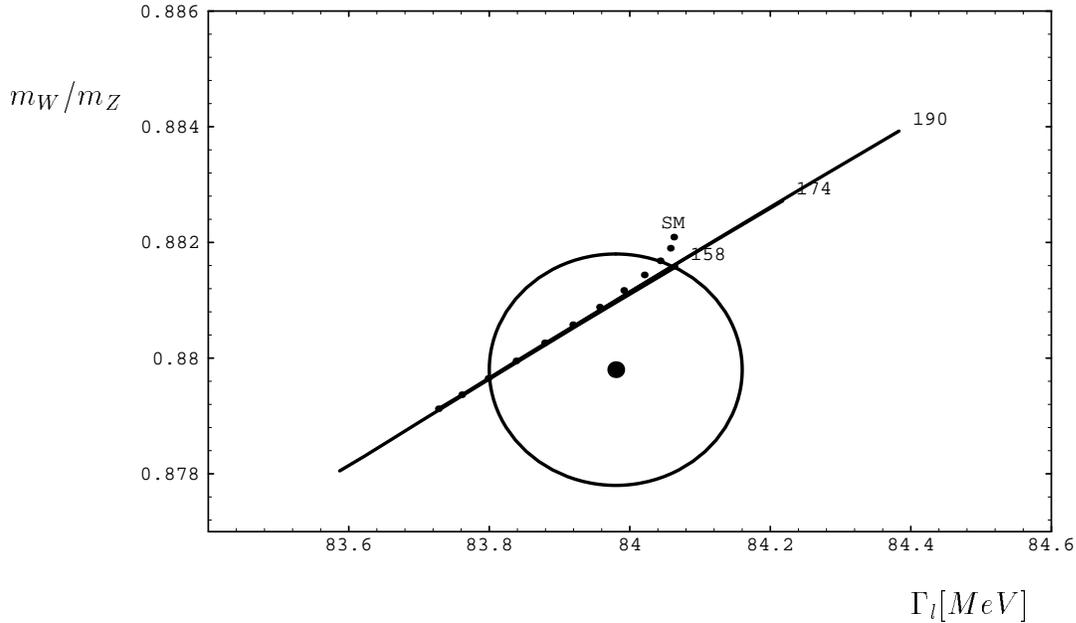

**Fig. 6.3** *The plot of $m_W/m_Z$ as the function of our "EW-meter" $\Gamma_l$ for top masses $m_t = 158 GeV$, $m_t = 174 GeV$ and $m_t = 190 GeV$. The dotted curve represents SM prediction for $m_t = 174 GeV$ and $m_H$ varying from 50GeV to 2.25TeV. The circle is plotted at the central experimental value and the ellipse is plotted at 1 standard deviation.*

Our model gives

$$(m_W/m_Z)^{HFM} = 0.8806 \pm 0.001.$$

This should be confronted with the experimental value [15]

$$(m_W/m_Z)^{EXP} = 0.8798 \pm 0.0020$$

and with SM prediction [2]

$$(m_W/m_Z)^{SM} = 0.8807 \pm 0.0002 \pm 0.0007.$$

We again see a reasonable agreement of our model predictions with experimental data and the coincidence of our predictions with the predictions of SM.

Finally we give in Fig. 6.4 the plot of our predictions for neutrino width of $Z$.

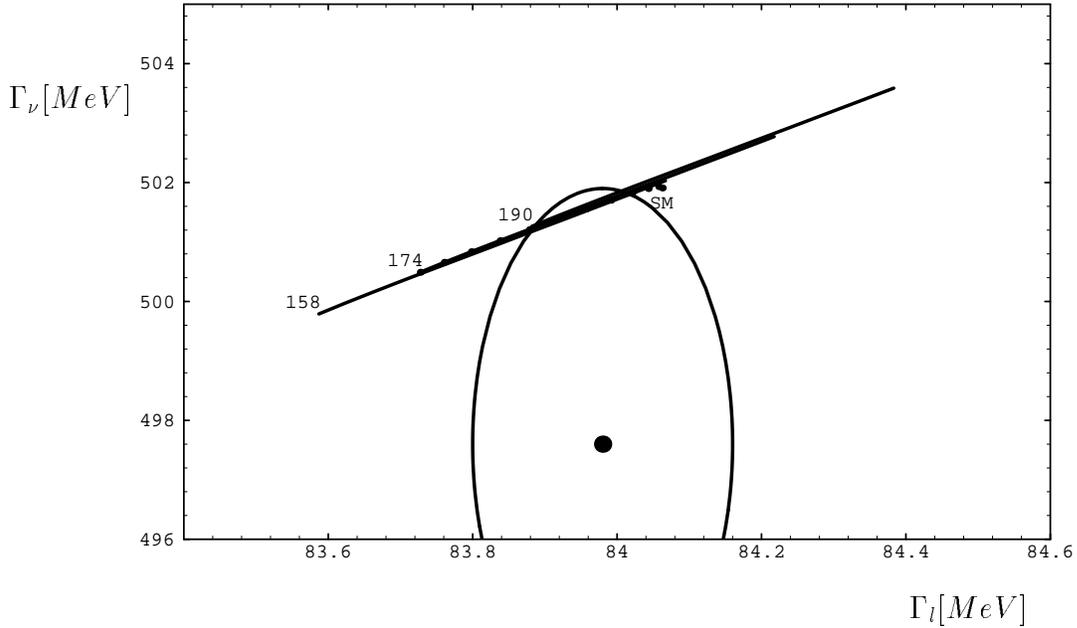

**Fig. 6.4** *The plot of $\Gamma_\nu$ as the function of our "EW-meter" $\Gamma_l$ for top masses $m_t = 158 GeV$, $m_t = 174 GeV$ and $m_t = 190 GeV$. The dotted curve represents SM prediction for $m_t = 174 GeV$ and $m_H$ varying from $50 GeV$ to $2.25 TeV$. The circle is plotted at the central experimental value and the ellipse is plotted at 1 standard deviation.*

Our model gives

$$(\Gamma_\nu)^{HFM} = 501.8 \pm 0.9 MeV.$$

This is consistent with the experimental data [3]

$$(\Gamma_\nu)^{EXP} = 497.6 \pm 4.3 MeV.$$

and, as one can see from Fig. 6.4, again coincides with the predictions of SM.

We can continue this review of predictions of our model and its confrontation with experimental data. We will observe complete agreement with the data influenced only by the low accuracy of measurements of some quantities. We will observe also that ours and SM predictions almost coincide what means that it will be extremely difficult to distinguish between these two models on the base of considered set of observables and within the present experimental accuracy.

The problem of elaboration of an effective calculational scheme for our model is considerably facilitated by the fact that introducing the suitable Stueckelberger auxiliary fields we can transform our model into the gauged nonlinear $\sigma$–model (GNL$\sigma$M) (see e.g. [20] and the discussion in Section 8). It is known that perturbative calculations in GNL$\sigma$M with cutoff $\Lambda$ are well elaborated and lead to interesting physical predictions for various processes[16],[20]. In addition it was recently shown that so called Generalized Equivalence Theorem (GET) holds in gauge field theories irrespectively if they are renormalizable or nonrenormalizable [37]. This remarkable theorem can be applied in the case of SM for heavy Higgs at high energy where

$$m_H, E \gg M_W, m_{f_i}$$

where E is the total energy and $m_{f_i}$ are lepton and quark masses respectively. It was shown that the leading parts coming from the L–loop diagrams are those diagrams for which N defined as

$$N = power\ of\ m_H + power\ of\ E \qquad (6.7)$$

becomes maximal. Using GET one relatively easily determines the leading contribution for any L–loop in SM and obtains high energy limit of a given scattering amplitude [37]. In the case of Higgs-free nonrenormalizable gauge field theory model one introduces cutoff $\Lambda$: in this case at high energy limit defined by inequalities

$$\Lambda > E \gg M_W, m_{f_i}$$

the leading diagrams are those for which

$$N = power\ of\ \Lambda + power\ of\ E \tag{6.8}$$

is maximal. Comparing (6.7) with (6.8) we see – as in the case of the $\varepsilon_{Ni}$-parameters – that the UV cutoff $\Lambda$ in Higgs-free gauge models replaces the large mass $m_H$. Using the criterion (6.8) and GET one obtains the high energy limit of scattering amplitude for various processes also in the non-renormalizable gauge models, like e.g. in the Higgs-free GNL$\sigma$M [37].

We see therefore that nonrenormalizability does not prevent us from getting definite predictions for physical processes in the low or high energy region from our model. Consequently the nonrenormalizable Higgs-free models may be as useful in a description of experimental data as the conventional SM.

## 7 Gravity Sector

Let us impose the scale fixing condition (4.13) on the lagrangian (4.2) and collect all gravitational terms. The lagrangian reads:

$$L^{scaled} = [L^{scaled}_{matter} - \frac{1}{12}(1+\beta)v^2 R - 2\rho(R^{\mu\nu}R_{\mu\nu} -$$

$$\frac{1}{3}R^2) - \frac{\lambda}{4}v^4]\sqrt{-g} \tag{7.1}$$

where we have selected the part $L^{scaled}_{matter}$ (describing the matter interacting with gravity) from the remaining purely gravitational terms.

The variation of (7.1) with respect to the metric $g^{\mu\nu}$ leads to the following classical equation of motion:

$$\rho[-\frac{2}{3}R_{;\mu;\nu} + 2R_{\mu\nu}{}^{;\eta}{}_{;\eta} - \frac{2}{3}g_{\mu\nu}R^{;\eta}{}_{;\eta} -$$

$$4R^{\eta\lambda}R_{\mu\eta\nu\lambda} + \frac{4}{3}RR_{\mu\nu} + g_{\mu\nu}(R^{\eta\lambda}R_{\eta\lambda} - \frac{1}{3}R^2)] +$$

$$\frac{1}{12}(1+\beta)v^2(R_{\mu\nu} - \frac{1}{2}g_{\mu\nu}R) + \frac{\lambda}{8}v^4 g_{\mu\nu} = \frac{1}{2}T_{\mu\nu}. \tag{7.2}$$

In the empty case $T_{\mu\nu} = 0$ this equation is satisfied by all solutions of an empty space Einstein equation with a properly chosen cosmological constant $\Lambda$:

$$R_{\mu\nu} - \frac{1}{2}g_{\mu\nu}R + \Lambda g_{\mu\nu} = 0. \tag{7.3}$$

In fact (7.3) implies that

$$R_{\mu\nu} \sim g_{\mu\nu} \quad \Rightarrow \quad R_{\mu\nu} = \frac{1}{4} R g_{\mu\nu} \tag{7.4}$$

and then

$$R_{\mu\nu} = \Lambda g_{\mu\nu}. \tag{7.5}$$

Inserting (7.4) into (7.2) we find that the part proportional to $\rho$ vanishes. The remnant can be collected leading to the relation

$$\frac{1}{8} v^2 g_{\mu\nu} \left(\frac{2}{3}(1+\beta)\Lambda - \lambda v^2\right) = 0 \tag{7.6}$$

where the empty space condition $T_{\mu\nu} = 0$ were used for the right hand side of (7.6).

Equation (7.6) implies

$$\Lambda = \frac{3}{2(1+\beta)} \lambda v^2. \tag{7.7}$$

Equation (7.7) relates the undetermined so far coupling constant $\lambda$ with a potentially observable cosmological constant $\Lambda$.

Let us go back to the case with the matter. Observe that the term linear in the curvature appears in (7.1) with the coefficient $-\frac{1}{12}(1+\beta)v^2$. If we want to reproduce the correct gravitational sector already at the classical level we have to admit for nonzero $\beta$ coupling. This would lead us to a model which is equivalent to the nonrenormalizable gauged nonlinear sigma model in the material sector. Accepting this price we can put

$$-\frac{1}{12}(1+\beta)v^2 = \kappa^{-2} \tag{7.8}$$

reproducing the Newtonian coupling in front of curvature $R$ in (7.1). This would mean that $\beta \approx -10^{38}$! Notice however that taking the scale fixing condition (4.13) the term $\beta \partial_\mu |\Phi| \partial^\mu |\Phi|$ vanishes. Hence it looks like that the only role of this term is to generate the proper value of Newton constant in the Einstein–Hilbert tree level lagrangian resulting from the Penrose term. (For further discussion see [10].)

The cosmological constant $\Lambda$ given by (7.7) was obtained from the analysis of gravitational interactions in the empty space-time. In reality the matter is always present and modifies the formula for $\Lambda$. In this case the most natural definition of effective cosmological constant was given by Zel'dovich [41] and by Adler [42] by means of the partition function determined by the lagrangian (5.1). (See also the excellent analysis of this problem in [43].)

# 8  Discussion.

The elementary particle physics is at present at a crossroad. We have in fact three drastically different alternatives:

$I^o$  The Higgs particle exists, its mass will be experimentally determined and will have the value predicted by the radiative corrections of SM. This will confirm the SSB mechanism for mass generation, the validity of SM framework and it will represent an extraordinary success of quantum gauge field theory.

$II^o$  The Higgs particle exists but its mass is considerable different from that predicted by the radiative corrections of SM. This would signal some kind of "New Physics" which will imply a reformulation of the present version of SM.

$III^o$  The Higgs particle does not exists. This will lead to a rejection of SM with Higgs sector and it will give preference to Higgs-free models for fundamental interactions. Presumably the obtained physical Higgs-free models will be nonrenormalizable.

It may be that the renormalizability of Quantum Gravity determined by Einstein–Hilbert action integral coupled with matter fields is not an "accident at work in quantum field theory" but it represents a universal feature that physical fundamental interactions considered simultaneously are nonrenormalizable. In this situation we are compulsed to use the nonrenormalizable models of quantum field theory for a description of fundamental interactions and we have to learn how to deduce predictions for experiments from such models.We have shown in Section 6 how to deduce the prediction for observables in our nonrenormalizable model. We have demonstrated that our predictions are in the surprising agreement with the experimental data. In fact the direct calculations of electroweak parameters $\varepsilon_{N1}$, $\varepsilon_{N2}$ and $\varepsilon_{N3}$ demonstrate that the Standard Model and the present model results differ effectively by the term proportional to $\log \frac{\Lambda^2}{m_H^2}$: thus it looks like that the very high Higgs mass $m_H$ plays in SM the role of the UV cutoff which in the present model may be replaced by parameter $\Lambda$. Thus the predictive power of our model may be comparable with that of the conventional SM.

In view of the possibility that nonrenormalizable nonabelian massive gauge field theories have to be used for a description of fundamental interactions it seems necessary to develop perturbative and nonperturbative methods for extracting predictions for scattering amplitudes and observables from such models [36]. In particular one should develop the corresponding Generalized Equivalence Theorems and determine explicitly the high energy behavior of cross sections in such models. The comparison of the obtained results with analytic formulas coming from Lipatov calculations [44] would be very inspiring. It would be also useful to develop systematic two–loop calculus with UV cutoff $\Lambda$ for electroweak processes. We plan in a near future to present several examples of such calculations.

The present model allows to obtain the Einsteinian form of gravitational interactions in the classical limit. It can be also analyzed by means of effective action for induced gravity [43].

ACKNOWLEDGMENTS

The authors are grateful to Prof. Iwo Białynicki Birula, Dr. B. Grządkowski, Prof. Z. Haba, Dr. M. Kalinowski and Prof. J. Werle for interesting discussions and Dr. S.D. Odintsov for sending to us the results of his group. They are especially grateful to Dr. S. Dittmaier for sending them his computer code and to Prof. D. Schildknecht for the extensive discussion of properties of his model.

# Appendix.

We prove here the Theorem 4.1. The measures in (4.17) have the form:

$$D\Phi = \prod_x d\Phi(x)$$

$$DA = \prod_{x,a,\mu,\nu} dA_\mu^a(x) dB_\nu(x) \qquad (A.1)$$

$$D\psi = \prod_{x,i} d\bar{\psi}_i(x) d\psi_i(x)$$

and according to [33]

$$Dg = \prod_{x,\mu \geq \nu} (-g(x))^{5/2} dg^{\mu\nu}(x) \qquad (A.2)$$

Let $\delta[g(\Psi)]$ be an another scale fixing condition. We show that the integral

$$Z^{-1} \int C(\Psi) e^{iS_T(\Psi)} \Delta_g(\Psi) \delta[g(\Psi)] D\Psi \qquad (A.3)$$

coincides with (4.15).

Note first that the measure $Dg$ is conformal invariant but the full measure $D\Psi$ given by (4.17) is not conformal invariant. It is crucial however that $D\Psi$ is multiplicative conformal covariant. In fact from (3.3) it follows that

$$D\Psi^\Omega = \rho(\Omega) D\Psi \qquad (A.4)$$

where $\Psi_i^\Omega = \Omega^{s_{\Psi_i}} \Psi_i$ is the conformal transform of $\Psi_i$, $\rho(\Omega) = \prod_x \Omega^{N_\Psi}(x)$ and $N_\Psi = \sum_i s_{\Psi_i}$ is the sum of conformal degrees of scalar, fermion and vector fields.

Then from (4.15), (4,17) and (A.4) we have

$$< C(\Psi) >_0 = Z^{-1} \int C(\Psi) e^{iS_T(\Psi)} \Delta_f(\Psi) \delta[f(\Psi^\Omega)] D\Psi D\Omega$$

$$= Z^{-1} \int C(\Psi) e^{iS_T(\Psi)} \Delta_f(\Psi) \delta[f(\Psi)] \Delta_g(\Psi) \delta[g(\Psi^{\Omega'})] D\Psi \rho^{-1}(\Omega) D\Omega D\Omega' \quad (A.5)$$

Setting $\Psi^{\Omega'} = \Psi'$ and using the multiplicative covariance of $D\Psi$ measure and the invariance of $D\Omega$ measure we obtain

$$\int \rho^{-1}(\Omega) \rho^{-1}(\Omega') D\Omega' = \int \rho^{-1}(\Omega \Omega') D\Omega' = c \quad (A.6)$$

The same constant appears in the partition function $Z$ and these constants cancel out in (A.5). Hence using the invariance of $D\Omega$ under inversion $\Omega \to \Omega^{-1}$ and (4.19) we obtain

$$< C(\Psi) >_0 = Z^{-1} \int C(\Psi) e^{iS_T(\Psi)} \Delta_g(\Psi) \delta[g(\Psi)] D\Psi. \quad (A.7)$$